\documentclass[a4paper,11pt]{article}\usepackage{jcappub}\def\PRDonly#1{}\def\JCAPonly#1{#1}

\usepackage{orcidlink,graphicx,cmap}
\usepackage[utf8]{inputenc}
\usepackage[T1]{fontenc}

\PRDonly{\newcommand{\Abstract}[1]{\begin{abstract}#1\end{abstract}}}
\JCAPonly{\newcommand{\Abstract}[1]{\abstract{#1}}}

\begin{document}

\title{Massive fields affected by echoes: New physics vs. astrophysical environment}

\PRDonly{%
\author{R. A. Konoplya \orcidlink{0000-0003-1343-9584}}
\email{roman.konoplya@gmail.com}
\affiliation{Research Centre for Theoretical Physics and Astrophysics, Institute of Physics, Silesian University in Opava, Bezručovo náměstí 13, CZ-74601 Opava, Czech Republic}
}%

\PRDonly{%
\author{Z. Stuchlík \orcidlink{0000-0003-2178-3588}}
\email{zdenek.stuchlik@physics.slu.cz}
\affiliation{Research Centre for Theoretical Physics and Astrophysics, Institute of Physics, Silesian University in Opava, Bezručovo náměstí 13, CZ-74601 Opava, Czech Republic}
}%

\PRDonly{%
\author{A. Zhidenko \orcidlink{0000-0001-6838-3309}}
\email{olexandr.zhydenko@ufabc.edu.br}
\affiliation{Centro de Matemática, Computação e Cognição (CMCC), Universidade Federal do ABC (UFABC), \\ Rua Abolição, CEP: 09210-180, Santo André, SP, Brazil}
}%

\JCAPonly{%
\author[\dagger]{Roman A. Konoplya \orcidlink{0000-0003-1343-9584},}
\emailAdd{roman.konoplya@gmail.com}
\author[\dagger]{Z. Stuchlík \orcidlink{0000-0003-2178-3588},}
\emailAdd{zdenek.stuchlik@physics.slu.cz}
\affiliation[\dagger]{Research Centre for Theoretical Physics and Astrophysics, \\ Institute of Physics, Silesian University in Opava, \\ Bezručovo náměstí 13, CZ-74601 Opava, Czech Republic}
}%

\JCAPonly{%
\author[\ddagger]{A. Zhidenko \orcidlink{0000-0001-6838-3309}}
\emailAdd{olexandr.zhydenko@ufabc.edu.br}
\affiliation[\ddagger]{Centro de Matemática, Computação e Cognição (CMCC), Universidade Federal do ABC (UFABC), \\ Rua Abolição, CEP: 09210-180, Santo André, SP, Brazil}
}%

\Abstract{
Unlike the perturbations of massless fields, the asymptotic tails of massive fields exhibit oscillations and decay slowly, following a power-law envelope. In this work, considering various scenarios admitting (either fundamental or effective) massive scalar and gravitational fields, we demonstrate that bump deformations in the effective potential, either in the near-horizon or far-field regions, modify these asymptotic oscillatory tails. Specifically, the power-law envelope transitions to a more complex oscillatory pattern, which cannot be easily fitted to a simple formula. This behavior is qualitatively different from the echoes of massless fields, which appear mainly during the quasinormal ringing stage and are considerably suppressed at the asymptotic tails. We show that in some models echoes may considerably amplify the signal at the stage of asymptotic tails.
}


\PRDonly{}

\JCAPonly{\arxivnumber{2411.09014}}

\maketitle

\section{Introduction}

Echoes are modifications of a signal at late times stipulated by the secondary scatterings of the wave from additional maxima of the effective potential \cite{Cardoso:2016oxy,Cardoso:2017cqb,Barausse:2014tra}. While echoes of massless fields have been studied in a great number publications (see, for instance, \cite{Cardoso:2016oxy,Cardoso:2017cqb,Barausse:2014tra,Huang:2021qwe,Konoplya:2018yrp,Abedi:2016hgu,Mark:2017dnq,Wang:2019rcf,Bronnikov:2019sbx,Bueno:2017hyj,Churilova:2021tgn,Guo:2022umh,Churilova:2019cyt,Li:2019kwa,Buoninfante:2020tfb,Liu:2020qia,Chowdhury:2022zqg,Yang:2024rms,Shen:2024rup,Yang:2024prm}
and references therein), echoes in the massive gravity, to the best of our knowledge, were discussed only in \cite{Dong:2020odp} where no effects for effectively massive fields were reported. However, the time-domain profiles for the Kaluza-Klein gravitational degrees of freedom in the braneworld scenario have been recently obtained in \cite{Tan:2024qij}.

At the same time, there are several motivations to study the evolution of perturbations in massive fields. Firstly, fields that are initially massless may acquire an effective mass term due to the influence of tidal forces in brane-world models and other extra-dimensional scenarios \cite{Seahra:2004fg,Ishihara:2008re}, or in the presence of a magnetic field \cite{Konoplya:2007yy,Konoplya:2008hj,Wu:2015fwa}. Additionally, massive gravitons and other massive particles are expected to contribute to long-wavelength gravitational waves \cite{Konoplya:2023fmh}, which are currently being investigated through Pulsar Timing Array observations \cite{NANOGrav:2023hvm}.

Although the LIGO detection of gravitational-wave signals has placed stringent constraints on the graviton mass $\mu$, or, equivalently, a lower bound on the graviton Compton wavelength $\lambda_c$ \cite{LIGOScientific:2016lio,LIGOScientific:2020tif},
\begin{equation}
    \mu c^2 \lesssim 2\cdot 10^{-23} eV, \quad \lambda_c \gtrsim 1~\mbox{light year},
\end{equation}
these limits apply primarily to theories in which gravity is mediated solely by a single massive spin-2 field.
In contrast, a broad class of alternative gravity theories admits both massless and massive gravitational degrees of freedom, with the massive modes typically arising in addition to the standard massless graviton. In such frameworks, the above observational bounds do not necessarily apply, and the massive sector can remain essentially unconstrained (see, e.g., the discussion in \cite{Konoplya:2025afm}).

Massive gravitational degrees of freedom, much like other massive fields, can also affect the evolution of gravitational perturbations through the nonlinear backreaction of the stress–energy carried by the waves, similarly to the gravitational-wave memory in massless gravity \cite{Heisenberg:2023prj}. A simplified setting of static perturbations of the massive field has been considered in \cite{Dong:2020odp}.

The quasinormal modes of massive fields differ qualitatively from those of massless fields, particularly due to the strong suppression of the damping rate. In some cases, this leads to the existence of arbitrarily long-lived modes, known as quasi-resonances, within the spectrum \cite{Ohashi:2004wr,Konoplya:2004wg,Konoplya:2006br,Zhidenko:2006rs,Bolokhov:2023bwm,Bolokhov:2023ruj,Lutfuoglu:2025hwh,Dubinsky:2025bvf}. Moreover, the asymptotic tails of massive fields exhibit distinct behavior, as they do not decay according to the power-law characteristic of massless fields. Instead, they oscillate and decay slowly with a power-law envelope (see the review in \cite{Konoplya:2011qq}).

Detecting these asymptotic tails poses substantial observational challenges: By the time the tail phase begins, signal amplitudes lie many orders of magnitude below the preceding quasinormal ringing and are deeply submerged in detector noise. For massless fields, the decay follows a smooth power-law falloff, quickly rendering the tail undetectable. However, for massive fields, the situation changes significantly: These tails not only decay more slowly than their massless counterparts, but can persist for much longer timescales, extending the observation window in which they might be extracted from noise. The oscillation frequency is determined by the black hole mass: For stellar-mass black holes, the frequency lies within the sensitivity band of ground-based detectors such as Advanced LIGO \cite{LIGOScientific:2014pky}. For supermassive black holes, the range shifts to the planned bandwidth of space-based antennas such as eLISA \cite{LISAConsortiumWaveformWorkingGroup:2023arg}. Although their amplitude is typically about $10^4$ times smaller than that of the corresponding quasinormal modes, the overlap between the natural frequency of the massive tail and the detector response, combined with the longer-lived nature of the signal, significantly improves the prospects for detection, provided that search algorithms can accommodate the distinctive oscillatory phase structure \cite{Degollado:2014vsa}.

In this work, we address a gap in the literature by studying echoes in massive fields. To this end, we consider three different models: the simplest massive scalar field in a Schwarzschild background, the squashed Kaluza-Klein black hole, and a wormhole mimicking the Schwarzschild solution \cite{Damour:2007ap}. In the first two models, echoes are induced by a deformation in  the effective potential, creating a bump. In contrast, the third model features a double-well potential, producing echoes without the need for an additional bump.

We have demonstrated that echoes in massive fields are markedly different from those in massless fields.
Specifically, when the bump is localised in the far zone, they affect the signal more during the asymptotic tail stage, where it induces oscillations in the power-law decay envelope.
Such deformations of the effective potential are typically associated with environmental effects, including the presence of accretion disks, nearby companion compact objects, active galactic nuclei, or extended matter distributions such as scalar or dark matter clouds (see, e.g., \cite{Konoplya:2018yrp} and references therein). The near-horizon bumps, one the contrary, induce the distinctive echoes effects in some cases at the earlier stage of quasinormal ringing. Given the slow decay of the asymptotic tails in massive fields, this enhances the significance of echoes in such fields.

The paper is organized as follows. In Section~\ref{sec:bump}, we examine the echoes of a massive scalar field in a Schwarzschild background, focusing on cases where the effective potential has a bump both at a distance and near the horizon. In Section~\ref{sec:KK}, we analyze the echoes of gravitational perturbations (effectively massive) around squashed Kaluza-Klein black holes. Section~\ref{sec:wormholes} is dedicated to the study of echoes of massive fields around the Schwarzschild-like wormhole.

\section{Schwarzschild black hole with the Gaussian bump in the effective potential}\label{sec:bump}

The simplest model for the decay of a massive field in a black hole background involves a minimally coupled massive scalar field in the Schwarzschild background. Echoes are then produced by a deformation in the effective potential's bump. A more self-consistent approach would require the deformation of the mass function to incorporate an appropriate bump. However, when the bump is either small or located far from the black hole, its back-reaction on the background metric can be safely neglected.

We consider the Schwarzschild black hole, which is given by the following line element
\begin{equation}\label{metric}
ds^2=-f(r)dt^2+\frac{dr^2}{f(r)}+r^2(d\theta^2+\sin^2\theta d\phi^2)\,,
\end{equation}
where
$$f(r)\equiv1-\frac{r_0}{r},$$
and $r_0=2GM$ is the radius of the event horizon.

The Klein-Gordon equation for a massive scalar field
\begin{equation}
    \frac{1}{\sqrt{-g}}\partial_\mu \left(\sqrt{-g}g^{\mu \nu}\partial_\nu\Phi\right)=\mu^2\Phi,
\end{equation}
after separation of the angular part
$$\Phi=\frac{1}{r}\Psi(t,r)Y_\ell^m(\theta,\phi),$$
can be reduced to the wavelike form
\begin{equation}\label{eq:wavelike}
\left(\frac{\partial^2}{\partial t^2}-\frac{\partial^2}{\partial r_*^2}+V(r)\right)\Psi(t,r)=0\,.
\end{equation}
Here $r_*$ is the tortoise coordinate,
$$dr_*=\frac{dr}{f(r)}\,,$$
and the effective potential is given by
\begin{equation}
V(r) = f(r)\left(\mu^2+\frac{\ell(\ell+1)}{r^2}+\frac{f'(r)}{r}\right)\,.
\end{equation}

We add a bump to the effective potential
\begin{equation}
V(r) \rightarrow V(r) + \delta V(r),
\end{equation}
where
\begin{equation}
    \delta V(r)=Af(r)e^{-(r-r_m)^2/\kappa}.
\end{equation}
Here $A$ is the bump size, $r_m$ is the point of its maximum and $\kappa$ is the width parameter.

To study the time-domain evolution of perturbations in these black hole models, we use the well-known time-domain integration scheme proposed in \cite{Gundlach:1993tp}.
We apply the discretization scheme in terms of the light-cone variables $u=t-r_*$ and $v=t+r_*$,
\begin{eqnarray}\label{Discretization}
\Psi\left(N\right)&=&\Psi\left(W\right)+\Psi\left(E\right)-\Psi\left(S\right)
\PRDonly{\\\nonumber&&}
-\Delta^2V\left(S\right)\frac{\Psi\left(W\right)+\Psi\left(E\right)}{8}+{\cal O}\left(\Delta^4\right)\,,
\end{eqnarray}
where we used the following notation for the points:
$N\equiv\left(u+\Delta,v+\Delta\right)$, $W\equiv\left(u+\Delta,v\right)$, $E\equiv\left(u,v+\Delta\right)$, and $S\equiv\left(u,v\right)$.
This discretization scheme has been widely applied in numerous studies (see \cite{Konoplya:2024ptj, Dubinsky:2024hmn, Dubinsky:2024jqi, Konoplya:2020bxa, Skvortsova:2024atk, Bolokhov:2023dxq, Bolokhov:2023ruj, Malik:2024tuf, Malik:2024sxv, Konoplya:2024lch} for recent examples).
The Gaussian initial data are imposed on the two null surfaces, $u=0$ and $v=0$:
\begin{equation}
\begin{array}{rcl}
    \Psi(u,0)&=&\exp\left(\dfrac{u-p}{2}\right)^2,\\
    \Psi(0,v)&=&\exp\left(\dfrac{v+p}{2}\right)^2,\\
\end{array}
\end{equation}
so that the positive value of the parameter $p$ corresponds to the Gaussian peak on the ``west'' null surface $v=0$ with respect to the observer, i.e., between the observer and the event horizon. The observer's position is fixed at $r_*=0$, which corresponds to the radial coordinate $r=r_p$.

Before discussing the effect of bumps onto the late-time decay of massive fields, let us first review the decay properties \emph{without} such bumps. The behavior of the late-time field can be divided into two main regimes.

At \emph{intermediate late times}, after the quasinormal ringing has decayed but before the asymptotic tail dominates, the field decays as a power law:
\begin{equation}
\Phi(t, r) \propto t^{-p} \sin(\mu t+\varphi),
\end{equation}
where $\varphi$ is a phase shift and the power index $p$ depends on the multipole number $\ell$ and spacetime properties. In the Schwarzschild background, for an observer at a fixed radius:
\begin{equation}
p = \ell + \frac{3}{2}.
\end{equation}
This behavior was first derived analytically in~\cite{Koyama:2001ee,Koyama:2001qw,Burko:2004jn} for Schwarzschild and Reissner-Nordström black holes.

At very late times ($\mu t \gg 1/\mu^2 M^2$), the field exhibits an oscillatory inverse power-law tail, commonly known as the \emph{asymptotic tail}:
\begin{equation}\label{asymptotictail}
\Phi(t, r) \propto t^{-5/6} \sin(\mu t + \varphi).
\end{equation}

This decay rate is \emph{universal}, in the sense that it does not depend on the multipole number $\ell$ or spacetime curvature, and it arises from backscattering off the asymptotically flat region. The power $-5/6$ was analytically derived in~\cite{Koyama:2001qw} and confirmed numerically in many follow-up works \cite{Konoplya:2025afm}. It is valid for Schwarzschild and other asymptotically flat black holes (such as Reissner-Nordström and slowly rotating Kerr).

Now, keeping in mind the qualitatively different late-time decay of massive fields, we are ready to discuss how potential bumps affect the evolution of perturbations. We can observe how the perturbations evolve over time, depending on various parameters such as the bump characteristics, the field's mass, and the observer's position. The modification of the late-time signal (echoes) originates from the partial reflection of the wave by a secondary peak in the effective potential, followed by its subsequent interaction with the black hole’s near-horizon region. Physically, this process corresponds to a portion of the outgoing radiation being trapped between the primary (near-horizon) barrier and the secondary, more distant potential barrier. Each round trip between these barriers generates a delayed, weaker replica of the original signal, giving rise to a sequence of echoes in the waveform. For a massless field, the characteristic delay time marking the onset of the first echo is approximately given by the light travel time from the black hole to the secondary peak and back. Therefore, this delay is approximately equal to twice the separation $\Delta r_*$ between the two potential peaks \cite{Cardoso:2016rao},
\begin{equation}\label{echotime}
    \Delta t\simeq2\Delta r_*=2\intop_{3r_0/2}^{~r_m}\frac{dr}{f(r)}.
\end{equation}
For a massive field, the modification of the late-time behavior also emerges on a similar timescale. In the regime of large field masses, the centrifugal barrier is suppressed, and the potential effectively resembles a step-like structure with a single bump rather than two distinct peaks. In this case, the modification of the late-time tail manifests as amplitude beating. Unlike the regular modulations arising from interference between long-lived quasinormal modes, such as those appearing due to superradiance \cite{Witek:2012tr}, this beating is irregular. As time progresses, the period of the beating gradually increases, leading to a sequence of increasingly prolonged wiggles in the amplitude. This behavior reflects the dispersive nature of the massive field and the evolving phase difference between overlapping wave components.

\begin{figure*}
\resizebox{\linewidth}{!}{\includegraphics{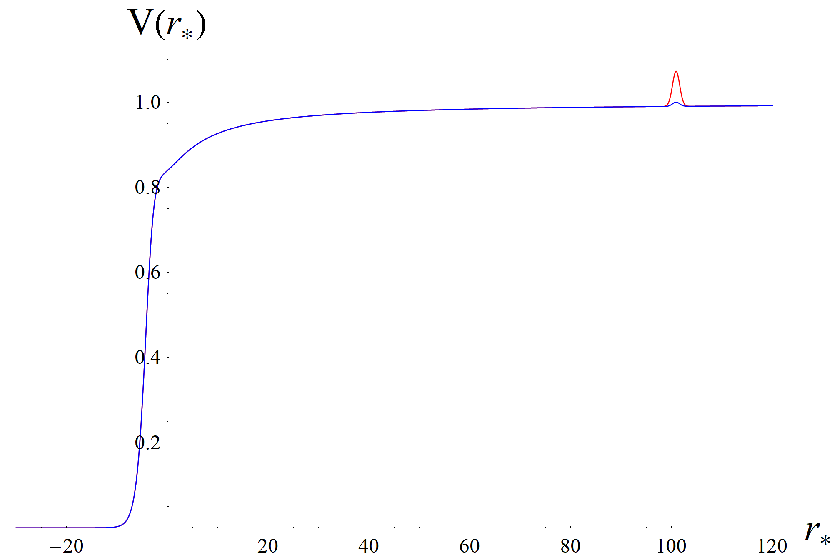}\includegraphics{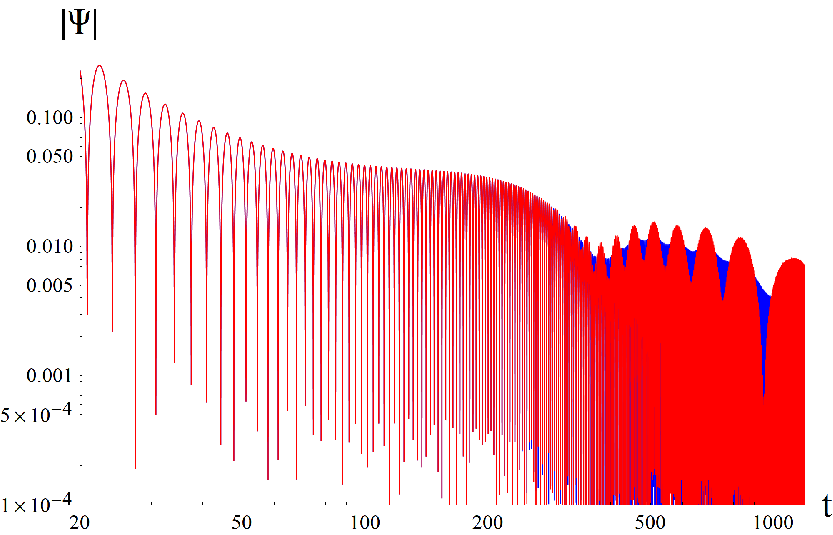}}
\caption{Effective potential and time-domain profile for a massive scalar field perturbations around the Schwarzschild black hole with a bump: $r_0=1$, $\ell=1$, $r_{m}=100$, $\kappa =1$, $\mu=1$, $A=1/100$ (blue) and $A=1/12$ (red). The observer is situated at $r_p=3$, and the center of the Gaussian wave-package is between the observer and the event horizon, $p=16$. The bump changes the asymptotic late-time behavior only.}\label{fig:massivebump}
\end{figure*}

\begin{figure*}
\resizebox{\linewidth}{!}{\includegraphics{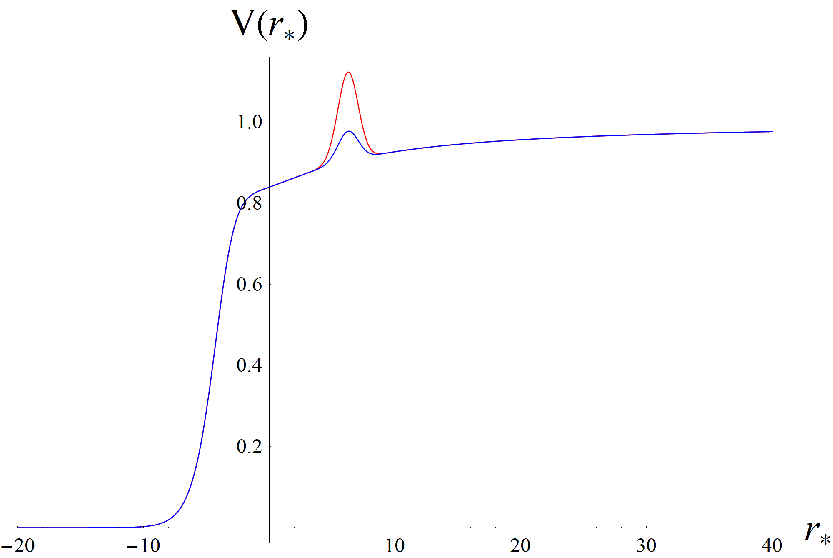}\includegraphics{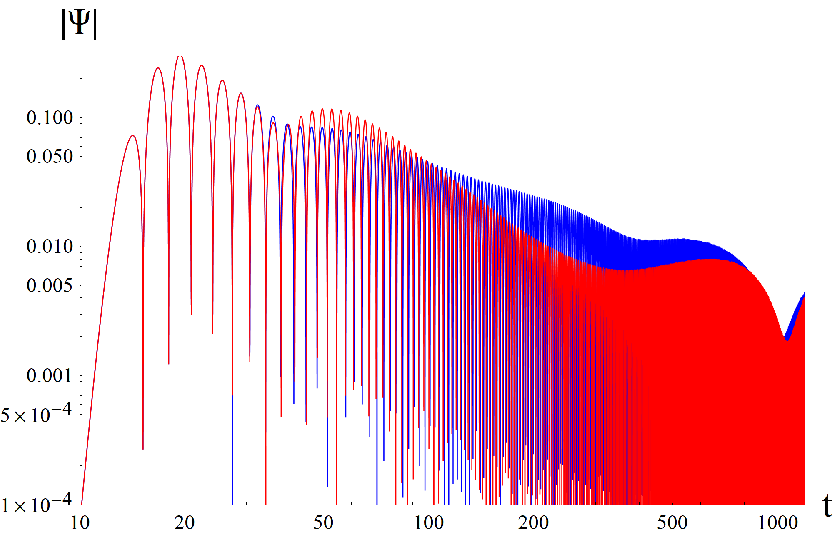}}
\caption{Effective potential and time-domain profile for a massive scalar field perturbations around the Schwarzschild black hole with a bump: $r_0=1$, $\ell=1$, $r_{m}=8$, $\kappa =1$, $\mu=1$, $A=1/12$ (blue) and $A=1/4$ (red). The observer is situated at $r_p=3$, and the center of the Gaussian wave-package is between the observer and the event horizon, $p=16$. The bump changes intermediate tails, but not the asymptotic late-time behavior.}\label{fig:massiveNHbump}
\end{figure*}

From the plot~\ref{fig:massivebump} one can see that, when the bump is located far from the black hole, in agreement with (\ref{echotime}) echoes appear at very late times, where we observe the modification of the power-law tail envelope. When the bump is closer to the event horizon, as shown in Fig.~\ref{fig:massiveNHbump}, both the intermediate tails and the final stage of the ringing phase are significantly modified. However, the asymptotic late-time tails, being governed by the asymptotic behavior of the potential, change insignificantly. Therefore, in order to observe the distinct echoes for the massive field, we need a significant modification of the effective potential outside the radiation zone. Such a modification can appear at a distance, due to another compact object with a comparable mass, e.g., a black hole or a neutron star, or, alternatively, due to the effects of new physics near the horizon or instead of the horizon. The latter effect for a Schwarzschild wormhole we shall consider in Sec.~\ref{sec:wormholes}.

For sufficiently small bumps near the event horizon, the time-domain profile during the ringdown phase does not differ significantly from that of the spacetime without a bump. This result aligns with \cite{Konoplya:2022pbc,Konoplya:2023hqb}, where it was demonstrated that the fundamental mode typically changes only slightly when the potential is slightly deformed near the horizon, with stronger deviations seen in the first few overtones. These overtones decay quickly, and for massive fields, their contribution to the time-domain profile is suppressed because (a) the fundamental mode decays more slowly and (b) the tails of massive fields begin to dominate earlier, leaving a relatively short period for the ringing phase.

\begin{figure*}
\resizebox{\linewidth}{!}{\includegraphics{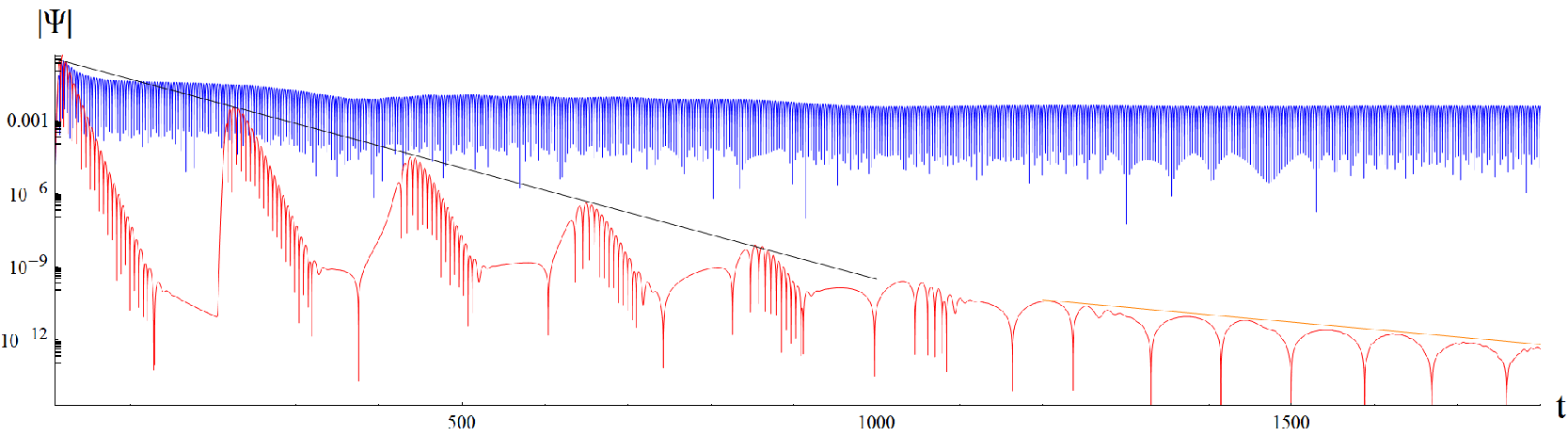}}
\resizebox{\linewidth}{!}{\includegraphics{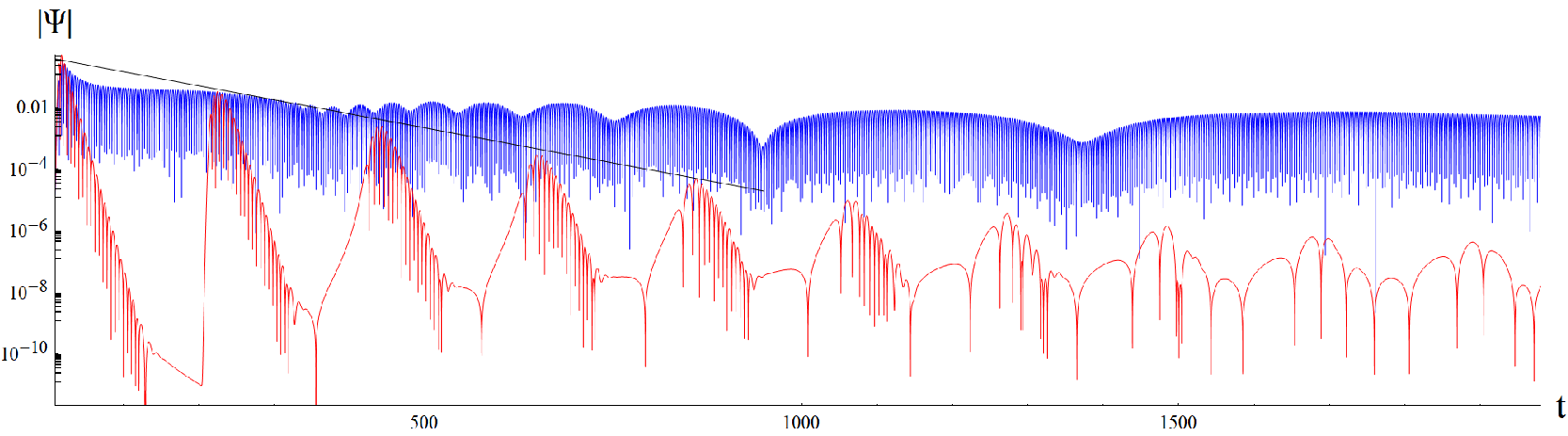}}
\caption{Semi-logarithmic plot of the time-domain profiles for a massive scalar field perturbations around the Schwarzschild black hole with a bump: $r_0=1$, $\ell=1$, $r_{m}=8$, $\kappa =1$, $\mu=1$ (blue) and $\mu=0$ (red), $A=1/100$ (top panel) and $A=1/12$ (bottom panel). The observer is situated at $r_p=3$, and the center of the Gaussian wave-package is between the observer and the event horizon, $p=16$. The black slopes show exponential decay of the echoes' amplitudes: $\propto\exp(-0.02t)$ for $A=1/100$ (top) and $\propto\exp(-0.01t)$ for $A=1/12$ (bottom). The orange slope in the top panel shows the final signal after relaxation, governed by the quasinormal modes of the composite potential. The higher the bump, the later the time at which full relaxation occurs.}\label{fig:massivemassless}
\end{figure*}

When comparing the echoes for massive and massless fields (see Fig.~\ref{fig:massivemassless}), we observe that the massless field, being long ranged, scatters the energy faster and, therefore the amplitude of echoes are many orders smaller. Even for an unrealistically, for astrophysical environment, large bump ($A=1/12$) we see that, although the amplitude of the first echo is of the same order of magnitude as the intermediate time tail, each next echo's amplitude becomes by orders smaller than the echo for the massive field. The amplitudes of the sequence of echoes decrease exponentially. Eventually, after the signal has relaxed, we observe an exponential decay governed by the quasinormal modes of the full potential characterized by two peaks.

\begin{figure*}
\resizebox{\linewidth}{!}{\includegraphics{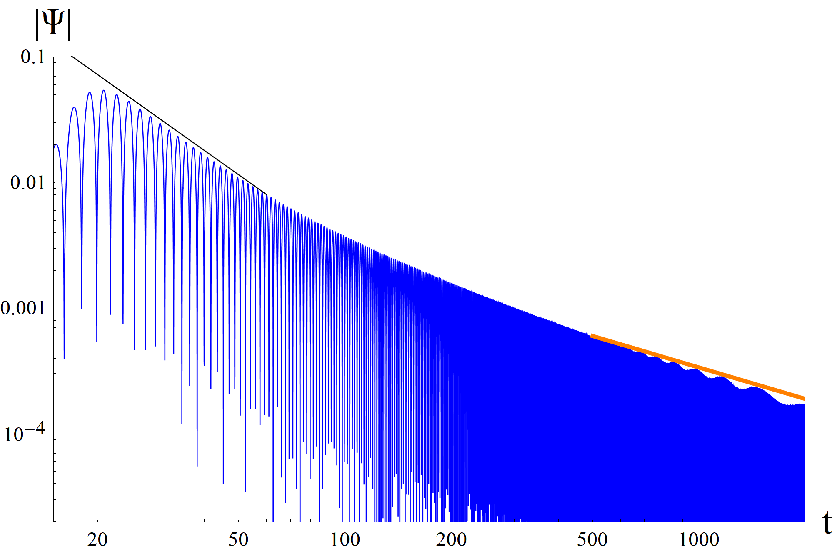}\includegraphics{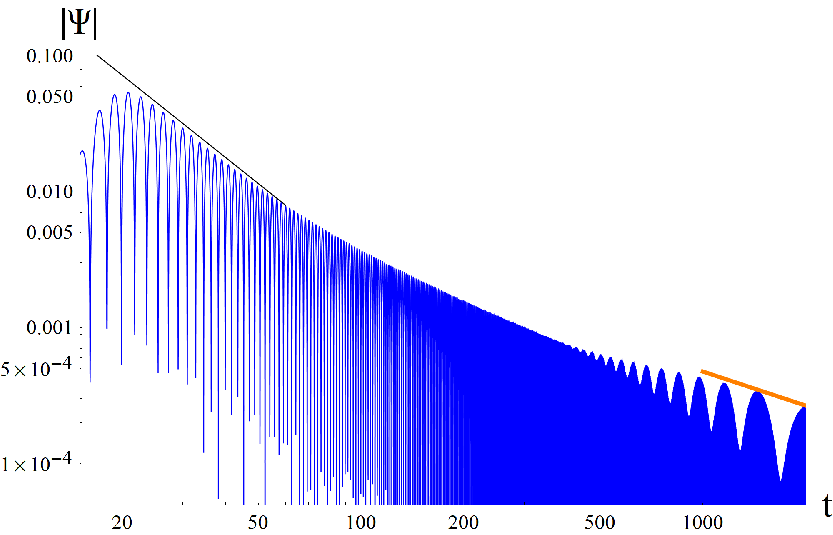}}
\caption{Logarithmic plot of the time-domain profiles for a massive scalar field perturbations around the Schwarzschild black hole with a bump: $r_0=1$, $\ell=1$, $r_{m}=8$, $\kappa =1$, $\mu=2$, $A=1/100$ (left panel) and $A=1/12$ (right panel). The black slopes show power-law decay of the intermediate tails ($\propto t^{-2}$) and the orange slopes show power-law decay of the amplitudes of the echoes at late times ($\propto t^{-5/6}$).}\label{fig:massivetails}
\end{figure*}

In contrast, a massive field decays according to a power-law tail. At intermediate times, the power-law tail interacts with the echoes generated by the bump, resulting in oscillations in the signal's amplitude. The period of these oscillations increases over time as the signal relaxes, eventually giving way to the universal asymptotic power-law tail decay for massive fields (see Fig.~\ref{fig:massivetails}), given by Eq.~(\ref{asymptotictail}).

Thus, perturbations of massive fields decay according to a power law at late times. As a result, the late-time amplitude of a massive field remains significantly higher (by several orders of magnitude) compared to that of a massless field. Even in the case of an unrealistically large bump, which enhances the amplitude of echoes, the exponential decay of massless fields continues into later times and ultimately results in a smaller amplitude compared to the late-time behavior of the massive field. This reflects the persistent power-law decay of the latter.

\section{Gravitational perturbations of squashed black holes in the Kaluza-Klein theory}\label{sec:KK}

Squashed Kaluza-Klein (KK) black holes \cite{Ishihara:2005dp} exhibit significant differences from standard Schwarzschild black holes, which are characterized by asymptotic flatness, and from black strings, even at energy levels where KK modes have not yet been activated. This makes squashed KK black holes a valuable model for exploring the dynamics of higher-dimensional spacetimes. Unlike their more conventional counterparts, these squashed black holes offer unique insights into higher dimensions, providing a glimpse into physics beyond the familiar four-dimensional framework.

One particularly noteworthy aspect of squashed KK black holes is their apparent stability \cite{Ishihara:2008re}. This distinguishes them from other higher-dimensional solutions, such as black strings, which are prone to the Gregory-Laflamme instability \cite{Gregory:1993vy}—a phenomenon that can cause certain configurations to become dynamically unstable and fragment. The inherent stability of squashed KK black holes allows us to bypass this instability, making them more viable candidates.

The metric of the uncharged non-rotating squashed Kaluza-Klein black hole is \cite{Ishihara:2008re}
\begin{eqnarray}
ds^2 &=&-F(\rho)dt^2 + \frac{G(\rho)^2}{F(\rho)}d\rho^2
\PRDonly{\\\nonumber&&}
+ \rho^2 G(\rho)^2 (\sigma_1^2+\sigma_2^2)
+\frac{\rho_{\infty}^2}{4 G(\rho)^2}(\sigma_3)^2,
\end{eqnarray}
where $\rho$ is the angular coordinate, and we employ the following angular basis:
\begin{eqnarray}
\nonumber
\sigma_1 &=& -\sin{\psi} d\theta+\cos{\psi} \sin{\theta}d\phi \ , \\
\sigma_2 &=& \cos{\psi} d\theta+\sin{\psi} \sin{\theta}d\phi \ , \\
\sigma_3 &=& d\psi+\cos{\theta}d\phi \ ,
\nonumber
\end{eqnarray}
and
$$
F(\rho) = 1-\frac{\rho_+}{\rho} \ , \qquad
G(\rho)^2 = 1+\frac{\rho_0}{\rho} \ .
$$
Here $\rho_+$ is the radius of the event horizon and $\rho_0$ defines the compact coordinate scale,
$$\rho_{\infty}^2 = 4 \rho_0(\rho_+ + \rho_0) \ ,$$
so that $\rho_0=\rho_{\infty}=0$ corresponds to the non-Kaluza-Klein (Schwarzschild) black hole.

As shown in \cite{Kimura:2007cr}, the linearized perturbation equations of the gravitational field for zero modes with respect to the $SU(2)$ group are also reduced to the wave-like form with the following effective potentials,
\begin{eqnarray}
V_0 &=& \frac{-(\rho_+-\rho)}{16\rho^3 (\rho+\rho_0)^3(4\rho+3\rho_0)^2}
\Big[256\rho_+\rho^4
\\\notag &&
+ 64\rho^3 (17\rho_++2\rho)\rho_0 +48\rho^2(32\rho_+ +11\rho)\rho_0^2
\PRDonly{\\\notag &&  \qquad}
+60\rho(13\rho_++12\rho)\rho_0^3+9(9\rho_+ +35\rho)\rho_0^4
\Big] \ ,
\\
V_1 &=& \left(1-\frac{\rho_+}{\rho}\right)
\bigg[\frac{1}{\rho_+\rho_0}
+\frac{7(-\rho_++\rho)}{16(\rho+\rho_0)^3}
\PRDonly{\\\notag &&}
+\frac{17 \rho_+-19\rho}{8 \rho(\rho+\rho_0)^2}
+\frac{9(-3\rho_++7\rho)}{16\rho^2(\rho+\rho_0)}
\JCAPonly{\\\notag &&}
+\frac{\rho_+-\rho}{\rho_+^2\rho +\rho_+\rho \rho_0}
\PRDonly{\\\notag &&}
-\frac{8(\rho_+-\rho)^2\rho}{(\rho_+-2\rho)(\rho_+\rho_0-\rho(\rho+2\rho_0))^2}
\PRDonly{\\\notag &&}
-\frac{2(\rho_+ +2\rho)}{(\rho_+-2\rho)(\rho_+\rho_0-\rho(\rho+2\rho_0))} \bigg] \ ,
\\
V_2 &=& \frac{ -(\rho_+-\rho)}{16\rho^3 \rho_0 (\rho_++\rho_0)(\rho+\rho_0)^3}
\Big[64\rho^5
\PRDonly{\\\notag &&}
+256\rho^4\rho_0-32\rho^3 \rho_0(\rho_+-11\rho_0)
\JCAPonly{\\\notag &&}
+9\rho_+ \rho_0^3(\rho_+ +\rho_0)
\PRDonly{\\\notag &&}
+8\rho^2\rho_0(2\rho_+^2 -5\rho_+\rho_0+25\rho_0^2)
\PRDonly{\\\notag &&}
+\rho \rho_0^2 (20\rho_+^2-9\rho_+\rho_0+35\rho_0^2)
\Big] \ .
\end{eqnarray}

Here $K=0,1,2$ are the eigenvalues corresponding to the $SU(1)$ group. $K=0$ is the
massless degree of freedom of the gravitational perturbations while $K=1$ and $K=2$ gain an effective mass.

\begin{figure*}
\resizebox{\linewidth}{!}{\includegraphics{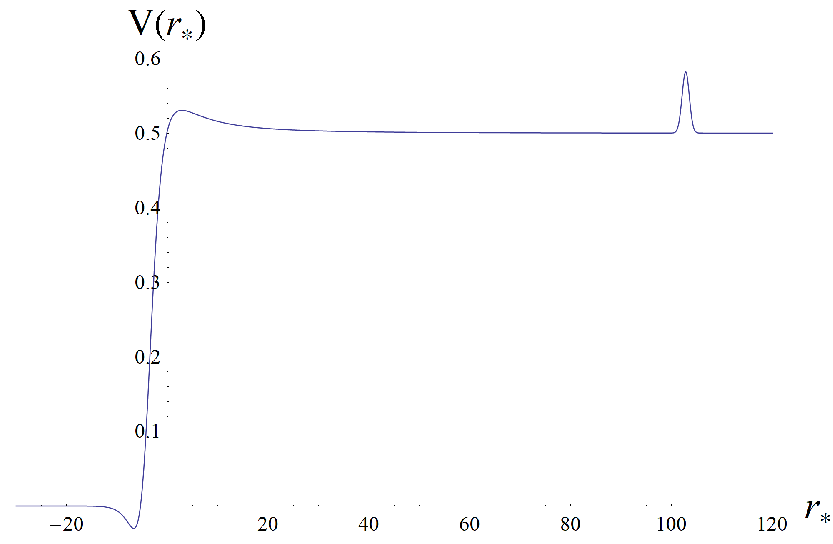}\includegraphics{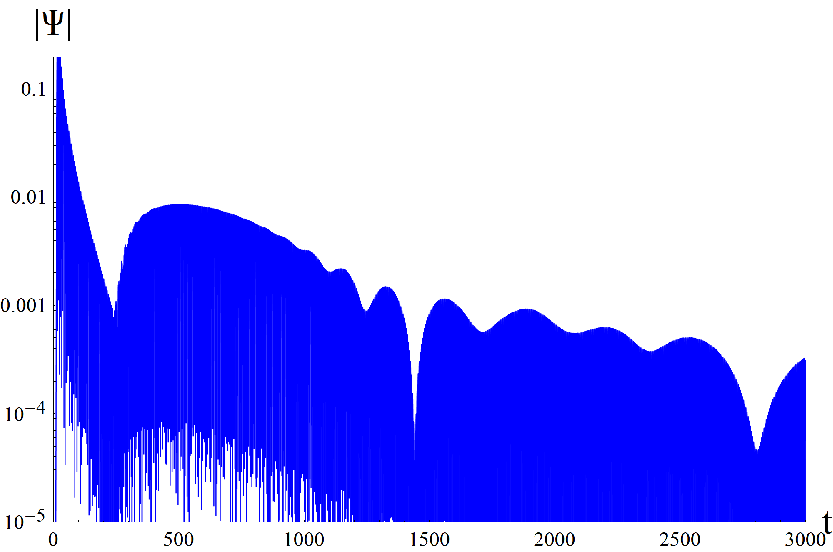}\includegraphics{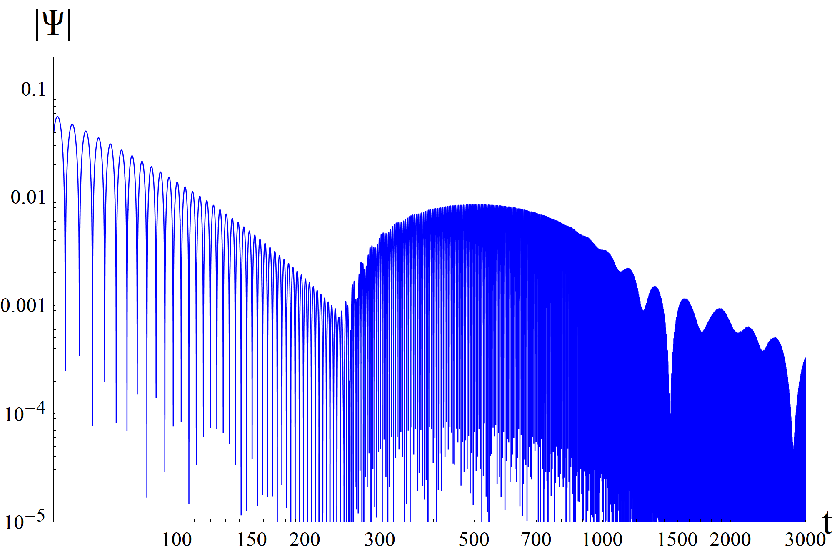}}
\resizebox{\linewidth}{!}{\includegraphics{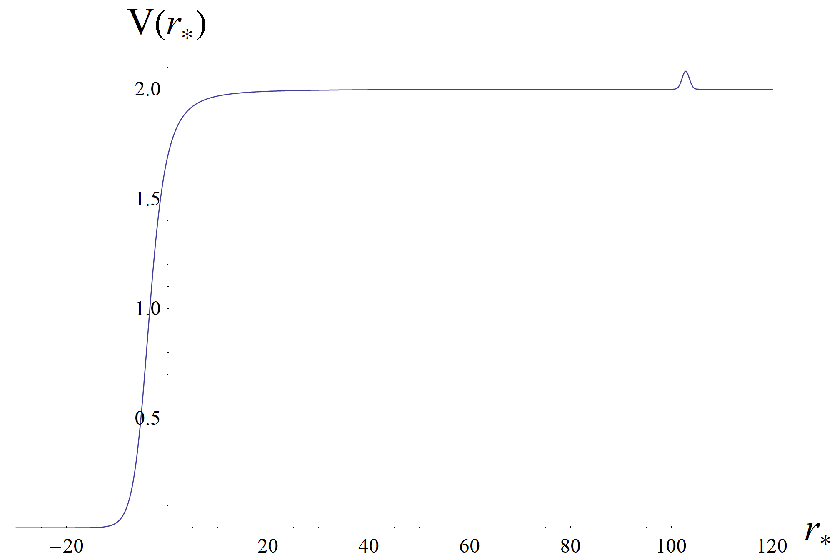}\includegraphics{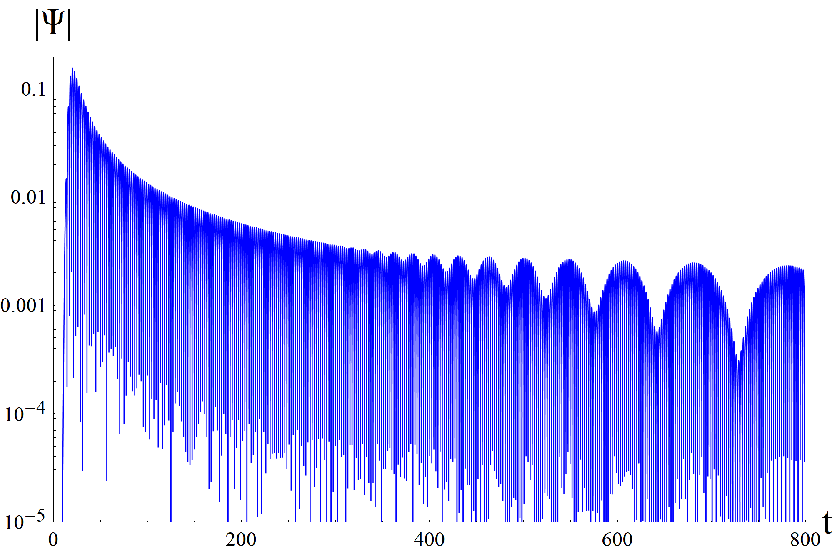}\includegraphics{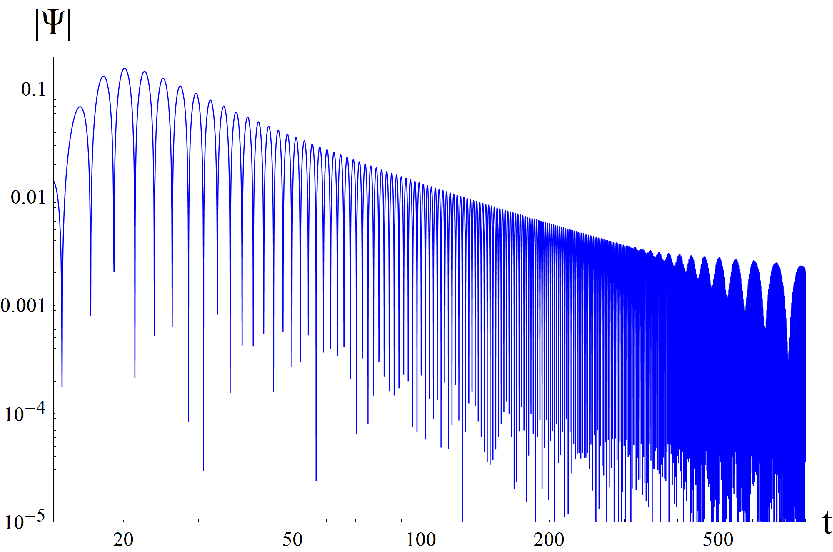}}
\caption{Effective potential and time-domain profile for $K=1$ (upper panels) $K=2$ (lower panels) perturbations around the squashed Kaluza-Klein black hole with a bump: $\rho_0=\rho_+=1$, $r_m=100$, $\kappa=1$, $A=1/12$. The central panels are for the semi-logarithmic plots, while the right panels are for the logarithmic plots. he observer is situated at $r_p=3$, and the center of the Gaussian wave-package is between the observer and the event horizon, $p=16$.}\label{fig:KKbump}
\end{figure*}

\begin{figure*}
\resizebox{\linewidth}{!}{\includegraphics{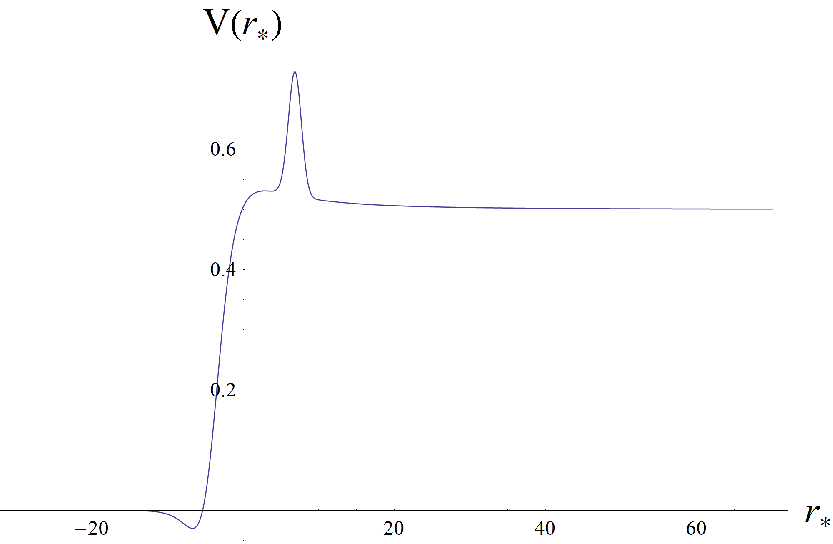}\includegraphics{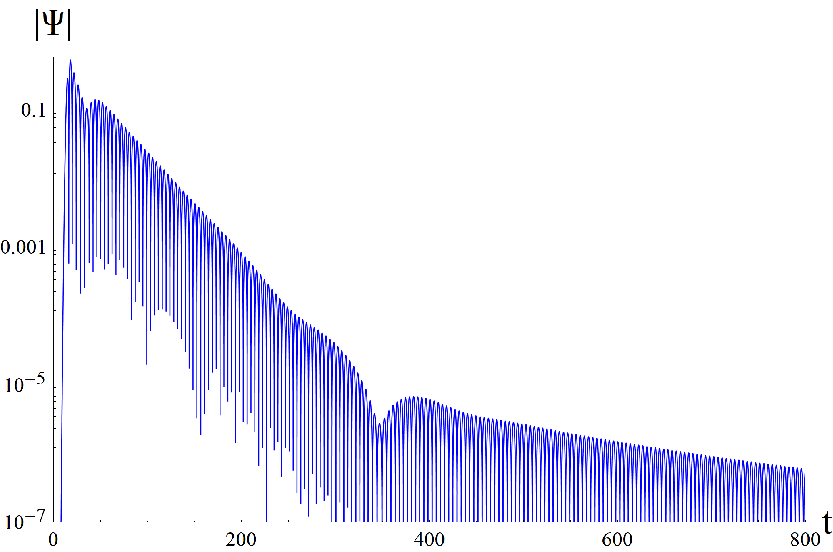}\includegraphics{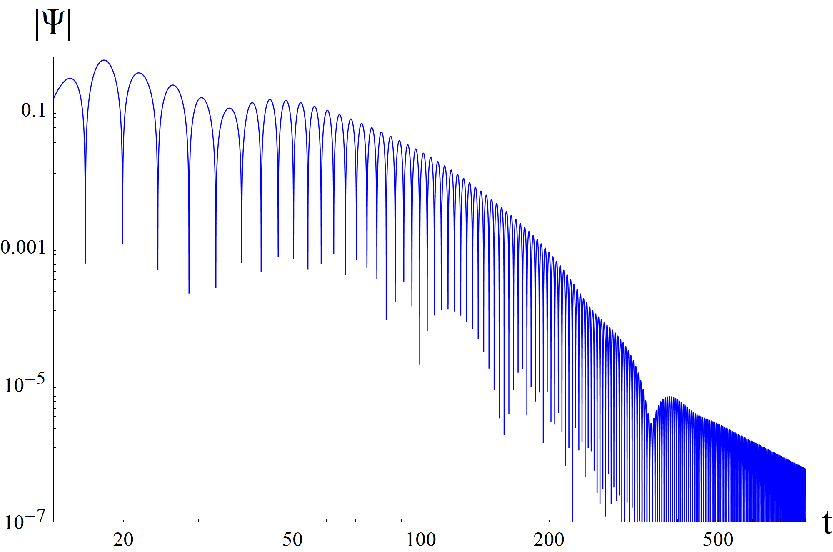}}
\resizebox{\linewidth}{!}{\includegraphics{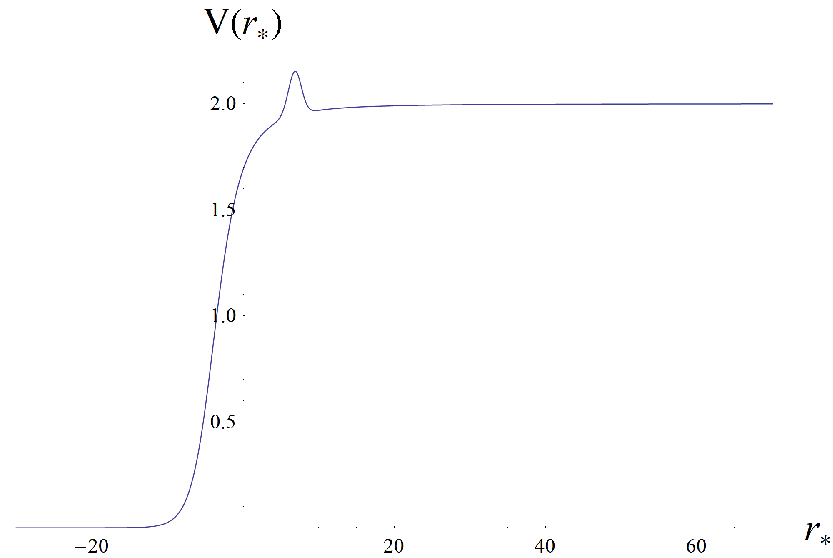}\includegraphics{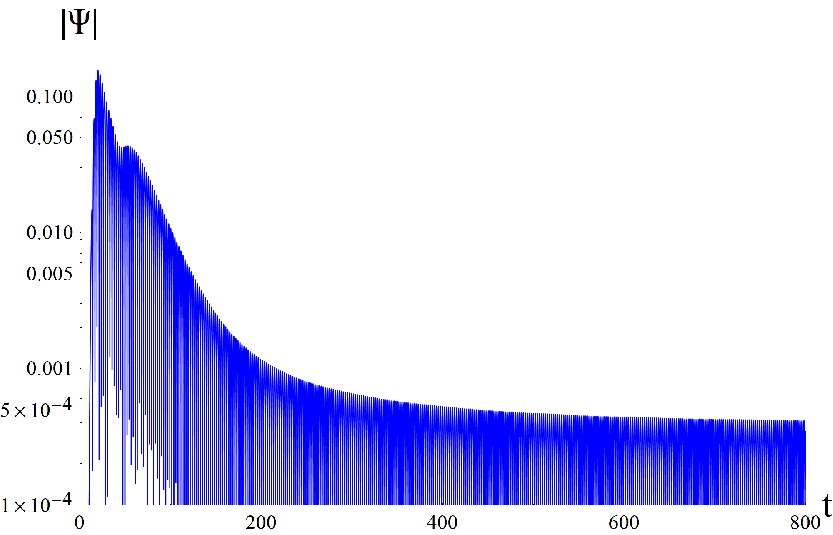}\includegraphics{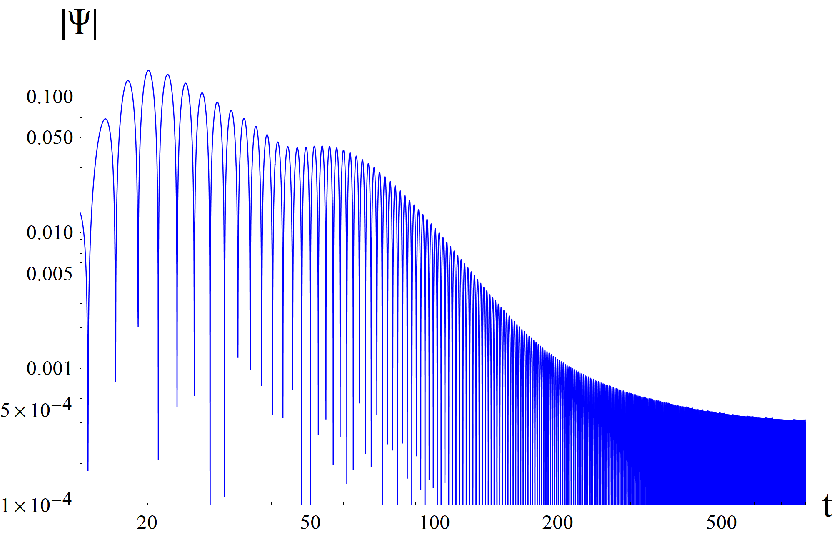}}
\caption{Effective potential and time-domain profile for $K=1$ (upper panels) $K=2$ (lower panels) perturbations around the squashed Kaluza-Klein black hole with a bump: $\rho_0=\rho_+=1$, $r_m=8$, $\kappa=1$, $A=0.25$. The central panels are for the semi-logarithmic plots, while the right panels are for the logarithmic plots. The observer is situated at $r_p=3$, and the center of the Gaussian wave-package is between the observer and the event horizon, $p=16$.}\label{fig:KKbumpNH}
\end{figure*}

From Figures \ref{fig:KKbump} and \ref{fig:KKbumpNH}, we observe that, similar to the case of massive scalar field perturbations, when the bump is located at a considerable distance from the black hole, the echoes start to modify the power-law behavior of the asymptotic tails, turning them into an oscillatory pattern. As the bump moves closer to the event horizon, the echoes appear earlier and begin to influence the ringing phase. Similar to the case of the Schwarzschild black hole, to observe echoes, the geometry needs to be modified at a distance from the radiation zone (Fig.~\ref{fig:KKbump}). A potential bump of the same size affects the $K=1$ perturbation profile more significantly due to the smaller effective mass.

\section{Schwarzschild-like wormholes}\label{sec:wormholes}

In this work, we consider $Z_2$-symmetric Schwarzschild-like wormholes \cite{Cardoso:2016rao} by assuming a Schwarzschild spacetime everywhere except at the wormhole throat, where a thin shell of matter is placed. The throat is positioned near the event horizon, at $r_t=r_0+\epsilon$ and we examine the region $r>r_t$  on both sides of the throat. For massless fields, the effective potential in this configuration takes the form of a double-peak potential, where the single peak in the Schwarzschild spacetime is split into two due to the symmetry of the radial coordinate. The echoes in this case arise from secondary scattering and reflection between the two peaks.

\begin{figure*}
\resizebox{\linewidth}{!}{\includegraphics{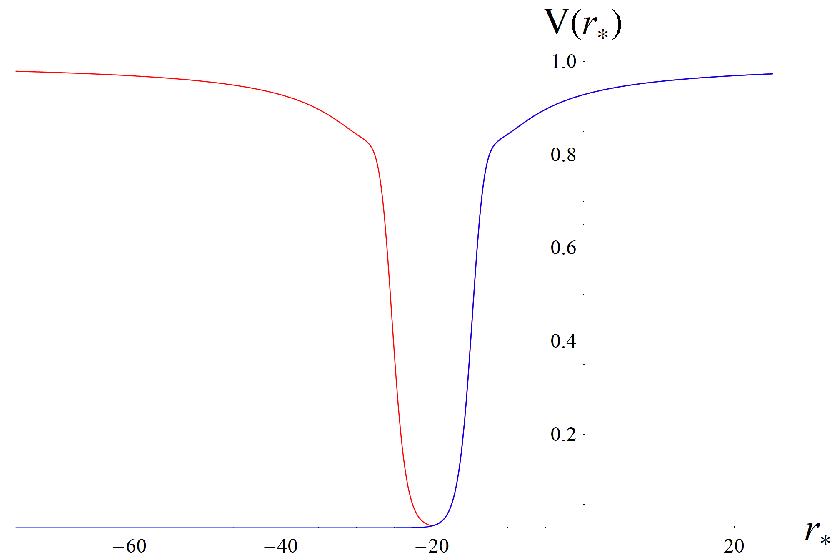}\includegraphics{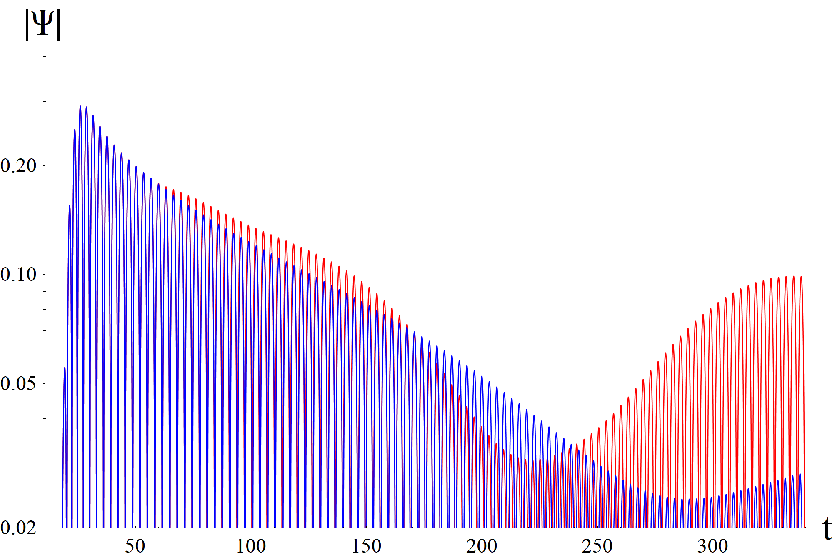}\includegraphics{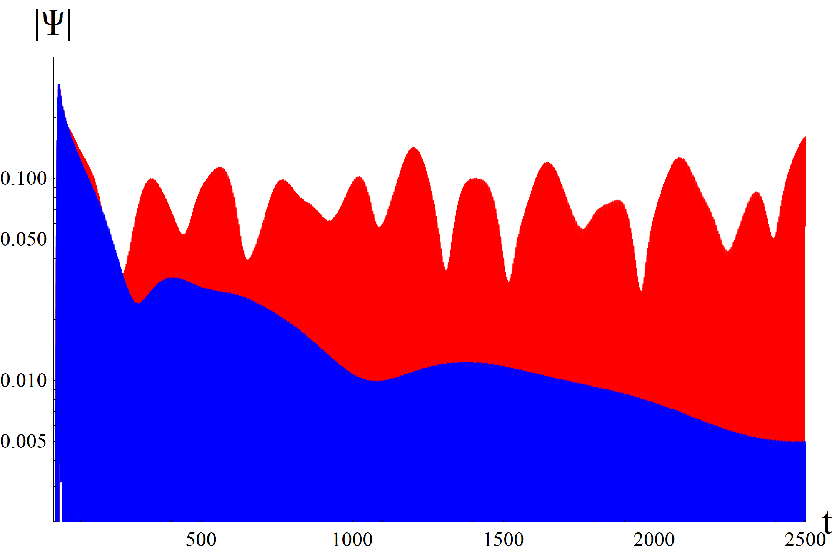}}
\caption{Effective potential (left) and time-domain profile for a massive scalar field perturbations around the Schwarzschild black hole (blue): $r_0=1$, $\ell=1$, $\mu=1$ and the wormhole with $r_t=1.001$ (red). The observer is situated at $r_p=11.721$, and the center of the Gaussian wave-package at the throat of the wormhole, $p=20$. The central panel shows the initial stage of the signal while the right panel shows also the asymptotically late-time behavior.}\label{fig:wormhole}
\end{figure*}

However, when a relatively large mass term is introduced, the potential shifts to a single-well form, approaching $\mu^2$ at both spatial infinities (see Figure \ref{fig:wormhole}). This modification results in distinct behavior in both the ringing phase and the asymptotic tails, while leaving the early stages of the ringing phase largely unaffected. The most notable effect is a significant amplification of the asymptotic tails, which reach a similar magnitude as the ringdown phase. Thus, the amplitudes of the massive-field echoes of the wormholes are larger than the echoes from the big bumps for the black holes (cf.~Fig.~\ref{fig:KKbump}). This demonstrates that echoes play a more prominent role for massive or effectively massive fields compared to massless ones, as massive fields decay more slowly, not only during the ringdown phase but also in the asymptotic tail stage.

\section{Conclusions}

Perturbations of massive fields evolve over time in a qualitatively different way from massless fields: after the relatively brief ringdown phase, asymptotic tails emerge, exhibiting an oscillatory pattern that decays slowly according to an exponential envelope law. This distinctive behavior of massive tails makes them promising candidates for the observation of long-wavelength gravitational waves, particularly through the Pulsar Timing Array experiment. This raises a natural question: could environmental factors or near-horizon effects (potentially arising from new physics) influence this radiation?

For massless fields, the answer is generally no—as long as the environment is localized at a significant distance from the black hole and its energy content is relatively low, as is the case with accretion disks around black holes. Mathematically, this stability arises because (a) the farther the center of the bump is, the later echoes appear, by which point the massless field has already decayed by several orders of magnitude; and (b) the smaller the bump, the weaker the echo amplitude.

In contrast, the slowly decaying and oscillatory tails of massive fields are significantly affected and amplified by echoes, which transform the power-law envelope into an oscillatory one. These amplified echoes remain comparable in magnitude to the signal's amplitude during the ringdown phase.

We also observed that a larger echo amplitude, resulting from a strong modification of the geometry either near the horizon or at a distance from the compact object, corresponds to a more irregular envelope shape at asymptotically late times. Additionally, the signal from massive fields decays more slowly due to the extra energy from the echoes, and, being comparable to the ringdown amplitude, may produce potentially observable effects.

There is one remark regarding our use of the term ``echoes''. In the massive case, when the mass is sufficiently large, the effective potential does not develop two distinct peaks. However, secondary scattering off the bump still occurs, leading to a late-time modulation of the signal, leading to the oscillatory power-law envelope. This behavior differs from the massless case, where the modulation appears as a sequence of ``ringdowns'' called echoes. We nevertheless adopt the term ``echoes'' here, as we are referring to modifications of the late-time signal caused by the presence of the bump.

Our work could be extended by including an analysis of the decay of massive fields in other scenarios that produce echoes, such as black hole/wormhole transitions in the brane-world model considered in \cite{Bronnikov:2019sbx,Bolokhov:2024voa} or various black hole models immersed in the galactic halo \cite{Konoplya:2025nqv}, either within specific models of the dark matter or an approximate solution for a black hole surrounded by a halo \cite{Konoplya:2022hbl}.

\JCAPonly{\bibliographystyle{JHEP}}

\bibliography{bibliography}

\providecommand{\href}[2]{#2}\begingroup\raggedright\begin{thebibliography}{10}

\bibitem{Cardoso:2016oxy}
V.~Cardoso, S.~Hopper, C.F.B.~Macedo, C.~Palenzuela and P.~Pani,
  \emph{{Gravitational-wave signatures of exotic compact objects and of quantum
  corrections at the horizon scale}},
  \href{https://doi.org/10.1103/PhysRevD.94.084031}{\emph{Phys. Rev. D}
  {\bfseries 94} (2016) 084031}
  [\href{https://arxiv.org/abs/1608.08637}{{\ttfamily 1608.08637}}].

\bibitem{Cardoso:2017cqb}
V.~Cardoso and P.~Pani, \emph{{Tests for the existence of black holes through
  gravitational wave echoes}},
  \href{https://doi.org/10.1038/s41550-017-0225-y}{\emph{Nature Astron.}
  {\bfseries 1} (2017) 586} [\href{https://arxiv.org/abs/1709.01525}{{\ttfamily
  1709.01525}}].

\bibitem{Barausse:2014tra}
E.~Barausse, V.~Cardoso and P.~Pani, \emph{{Can environmental effects spoil
  precision gravitational-wave astrophysics?}},
  \href{https://doi.org/10.1103/PhysRevD.89.104059}{\emph{Phys. Rev. D}
  {\bfseries 89} (2014) 104059}
  [\href{https://arxiv.org/abs/1404.7149}{{\ttfamily 1404.7149}}].

\bibitem{Huang:2021qwe}
H.~Huang, M.-Y.~Ou, M.-Y.~Lai and H.~Lu, \emph{{Echoes from classical black
  holes}}, \href{https://doi.org/10.1103/PhysRevD.105.104049}{\emph{Phys. Rev.
  D} {\bfseries 105} (2022) 104049}
  [\href{https://arxiv.org/abs/2112.14780}{{\ttfamily 2112.14780}}].

\bibitem{Konoplya:2018yrp}
R.A.~Konoplya, Z.~Stuchlík and A.~Zhidenko, \emph{{Echoes of compact objects:
  new physics near the surface and matter at a distance}},
  \href{https://doi.org/10.1103/PhysRevD.99.024007}{\emph{Phys. Rev. D}
  {\bfseries 99} (2019) 024007}
  [\href{https://arxiv.org/abs/1810.01295}{{\ttfamily 1810.01295}}].

\bibitem{Abedi:2016hgu}
J.~Abedi, H.~Dykaar and N.~Afshordi, \emph{{Echoes from the Abyss: Tentative
  evidence for Planck-scale structure at black hole horizons}},
  \href{https://doi.org/10.1103/PhysRevD.96.082004}{\emph{Phys. Rev. D}
  {\bfseries 96} (2017) 082004}
  [\href{https://arxiv.org/abs/1612.00266}{{\ttfamily 1612.00266}}].

\bibitem{Mark:2017dnq}
Z.~Mark, A.~Zimmerman, S.M.~Du and Y.~Chen, \emph{{A recipe for echoes from
  exotic compact objects}},
  \href{https://doi.org/10.1103/PhysRevD.96.084002}{\emph{Phys. Rev. D}
  {\bfseries 96} (2017) 084002}
  [\href{https://arxiv.org/abs/1706.06155}{{\ttfamily 1706.06155}}].

\bibitem{Wang:2019rcf}
Q.~Wang, N.~Oshita and N.~Afshordi, \emph{{Echoes from Quantum Black Holes}},
  \href{https://doi.org/10.1103/PhysRevD.101.024031}{\emph{Phys. Rev. D}
  {\bfseries 101} (2020) 024031}
  [\href{https://arxiv.org/abs/1905.00446}{{\ttfamily 1905.00446}}].

\bibitem{Bronnikov:2019sbx}
K.A.~Bronnikov and R.A.~Konoplya, \emph{{Echoes in brane worlds: ringing at a
  black hole--wormhole transition}},
  \href{https://doi.org/10.1103/PhysRevD.101.064004}{\emph{Phys. Rev. D}
  {\bfseries 101} (2020) 064004}
  [\href{https://arxiv.org/abs/1912.05315}{{\ttfamily 1912.05315}}].

\bibitem{Bueno:2017hyj}
P.~Bueno, P.A.~Cano, F.~Goelen, T.~Hertog and B.~Vercnocke, \emph{{Echoes of
  Kerr-like wormholes}},
  \href{https://doi.org/10.1103/PhysRevD.97.024040}{\emph{Phys. Rev. D}
  {\bfseries 97} (2018) 024040}
  [\href{https://arxiv.org/abs/1711.00391}{{\ttfamily 1711.00391}}].

\bibitem{Churilova:2021tgn}
M.S.~Churilova, R.A.~Konoplya, Z.~Stuchlik and A.~Zhidenko, \emph{{Wormholes
  without exotic matter: quasinormal modes, echoes and shadows}},
  \href{https://doi.org/10.1088/1475-7516/2021/10/010}{\emph{JCAP} {\bfseries
  10} (2021) 010} [\href{https://arxiv.org/abs/2107.05977}{{\ttfamily
  2107.05977}}].

\bibitem{Guo:2022umh}
G.~Guo, P.~Wang, H.~Wu and H.~Yang, \emph{{Echoes from hairy black holes}},
  \href{https://doi.org/10.1007/JHEP06(2022)073}{\emph{JHEP} {\bfseries 06}
  (2022) 073} [\href{https://arxiv.org/abs/2204.00982}{{\ttfamily
  2204.00982}}].

\bibitem{Churilova:2019cyt}
M.S.~Churilova and Z.~Stuchlik, \emph{{Ringing of the regular
  black-hole/wormhole transition}},
  \href{https://doi.org/10.1088/1361-6382/ab7717}{\emph{Class. Quant. Grav.}
  {\bfseries 37} (2020) 075014}
  [\href{https://arxiv.org/abs/1911.11823}{{\ttfamily 1911.11823}}].

\bibitem{Li:2019kwa}
Z.-P.~Li and Y.-S.~Piao, \emph{{Mixing of gravitational wave echoes}},
  \href{https://doi.org/10.1103/PhysRevD.100.044023}{\emph{Phys. Rev. D}
  {\bfseries 100} (2019) 044023}
  [\href{https://arxiv.org/abs/1904.05652}{{\ttfamily 1904.05652}}].

\bibitem{Buoninfante:2020tfb}
L.~Buoninfante, \emph{{Echoes from corpuscular black holes}},
  \href{https://doi.org/10.1088/1475-7516/2020/12/041}{\emph{JCAP} {\bfseries
  12} (2020) 041} [\href{https://arxiv.org/abs/2005.08426}{{\ttfamily
  2005.08426}}].

\bibitem{Liu:2020qia}
H.~Liu, P.~Liu, Y.~Liu, B.~Wang and J.-P.~Wu, \emph{{Echoes from phantom
  wormholes}}, \href{https://doi.org/10.1103/PhysRevD.103.024006}{\emph{Phys.
  Rev. D} {\bfseries 103} (2021) 024006}
  [\href{https://arxiv.org/abs/2007.09078}{{\ttfamily 2007.09078}}].

\bibitem{Chowdhury:2022zqg}
A.~Chowdhury, S.~Devi and S.~Chakrabarti, \emph{{Naked singularity in 4D
  Einstein-Gauss-Bonnet novel gravity: Echoes and instability}},
  \href{https://doi.org/10.1103/PhysRevD.106.024023}{\emph{Phys. Rev. D}
  {\bfseries 106} (2022) 024023}
  [\href{https://arxiv.org/abs/2202.13698}{{\ttfamily 2202.13698}}].

\bibitem{Yang:2024rms}
Z.-H.~Yang, C.~Xu, X.-M.~Kuang, B.~Wang and R.-H.~Yue, \emph{{Echoes of
  massless scalar field induced from hairy Schwarzschild black hole}},
  \href{https://doi.org/10.1016/j.physletb.2024.138688}{\emph{Phys. Lett. B}
  {\bfseries 853} (2024) 138688}.

\bibitem{Shen:2024rup}
S.-F.~Shen, K.~Lin, T.~Zhu, Y.-P.~Yan, C.-G.~Shao and W.-L.~Qian, \emph{{Two
  distinct types of echoes in compact objects}},
  \href{https://doi.org/10.1103/PhysRevD.110.084022}{\emph{Phys. Rev. D}
  {\bfseries 110} (2024) 084022}
  [\href{https://arxiv.org/abs/2408.00971}{{\ttfamily 2408.00971}}].

\bibitem{Yang:2024prm}
H.~Yang, Z.-W.~Xia and Y.-G.~Miao, \emph{{Echoes and quasi-normal modes of
  perturbations around Schwarzchild traversable wormholes}},
  \href{https://arxiv.org/abs/2406.00377}{{\ttfamily 2406.00377}}.

\bibitem{Dong:2020odp}
R.~Dong and D.~Stojkovic, \emph{{Gravitational wave echoes from black holes in
  massive gravity}},
  \href{https://doi.org/10.1103/PhysRevD.103.024058}{\emph{Phys. Rev. D}
  {\bfseries 103} (2021) 024058}
  [\href{https://arxiv.org/abs/2011.04032}{{\ttfamily 2011.04032}}].

\bibitem{Tan:2024qij}
Q.~Tan, S.~Long, W.~Deng and J.~Jing, \emph{{Quasinormal modes and echoes of a
  double braneworld}}, {\emph{arXiv:2410.06945} (2024) }
  [\href{https://arxiv.org/abs/2410.06945}{{\ttfamily 2410.06945}}].

\bibitem{Seahra:2004fg}
S.S.~Seahra, C.~Clarkson and R.~Maartens, \emph{{Detecting extra dimensions
  with gravity wave spectroscopy: the black string brane-world}},
  \href{https://doi.org/10.1103/PhysRevLett.94.121302}{\emph{Phys. Rev. Lett.}
  {\bfseries 94} (2005) 121302}
  [\href{https://arxiv.org/abs/gr-qc/0408032}{{\ttfamily gr-qc/0408032}}].

\bibitem{Ishihara:2008re}
H.~Ishihara, M.~Kimura, R.A.~Konoplya, K.~Murata, J.~Soda and A.~Zhidenko,
  \emph{{Evolution of perturbations of squashed Kaluza-Klein black holes:
  escape from instability}},
  \href{https://doi.org/10.1103/PhysRevD.77.084019}{\emph{Phys. Rev. D}
  {\bfseries 77} (2008) 084019}
  [\href{https://arxiv.org/abs/0802.0655}{{\ttfamily 0802.0655}}].

\bibitem{Konoplya:2007yy}
R.A.~Konoplya and R.D.B.~Fontana, \emph{{Quasinormal modes of black holes
  immersed in a strong magnetic field}},
  \href{https://doi.org/10.1016/j.physletb.2007.10.065}{\emph{Phys. Lett. B}
  {\bfseries 659} (2008) 375}
  [\href{https://arxiv.org/abs/0707.1156}{{\ttfamily 0707.1156}}].

\bibitem{Konoplya:2008hj}
R.A.~Konoplya, \emph{{Magnetic field creates strong superradiant instability}},
  \href{https://doi.org/10.1016/j.physletb.2008.11.059}{\emph{Phys. Lett. B}
  {\bfseries 666} (2008) 283}
  [\href{https://arxiv.org/abs/0801.0846}{{\ttfamily 0801.0846}}].

\bibitem{Wu:2015fwa}
C.~Wu and R.~Xu, \emph{{Decay of massive scalar field in a black hole
  background immersed in magnetic field}},
  \href{https://doi.org/10.1140/epjc/s10052-015-3632-1}{\emph{Eur. Phys. J. C}
  {\bfseries 75} (2015) 391}
  [\href{https://arxiv.org/abs/1507.04911}{{\ttfamily 1507.04911}}].

\bibitem{Konoplya:2023fmh}
R.A.~Konoplya and A.~Zhidenko, \emph{{Asymptotic tails of massive gravitons in
  light of pulsar timing array observations}},
  \href{https://doi.org/10.1016/j.physletb.2024.138685}{\emph{Phys. Lett. B}
  {\bfseries 853} (2024) 138685}
  [\href{https://arxiv.org/abs/2307.01110}{{\ttfamily 2307.01110}}].

\bibitem{NANOGrav:2023hvm}
{\scshape NANOGrav} collaboration, \emph{{The NANOGrav 15 yr Data Set: Search
  for Signals from New Physics}},
  \href{https://doi.org/10.3847/2041-8213/acdc91}{\emph{Astrophys. J. Lett.}
  {\bfseries 951} (2023) L11}
  [\href{https://arxiv.org/abs/2306.16219}{{\ttfamily 2306.16219}}].

\bibitem{LIGOScientific:2016lio}
{\scshape LIGO Scientific, Virgo} collaboration, \emph{{Tests of general
  relativity with GW150914}},
  \href{https://doi.org/10.1103/PhysRevLett.116.221101}{\emph{Phys. Rev. Lett.}
  {\bfseries 116} (2016) 221101}
  [\href{https://arxiv.org/abs/1602.03841}{{\ttfamily 1602.03841}}].

\bibitem{LIGOScientific:2020tif}
{\scshape LIGO Scientific, Virgo} collaboration, \emph{{Tests of general
  relativity with binary black holes from the second LIGO-Virgo
  gravitational-wave transient catalog}},
  \href{https://doi.org/10.1103/PhysRevD.103.122002}{\emph{Phys. Rev. D}
  {\bfseries 103} (2021) 122002}
  [\href{https://arxiv.org/abs/2010.14529}{{\ttfamily 2010.14529}}].

\bibitem{Konoplya:2025afm}
R.A.~Konoplya, A.~Spina and A.~Zhidenko, \emph{{Time evolution of black hole
  perturbations in quadratic gravity}},
  \href{https://doi.org/10.1103/xhtc-9cf4}{\emph{Phys. Rev. D} {\bfseries 112}
  (2025) 024060} [\href{https://arxiv.org/abs/2505.01128}{{\ttfamily
  2505.01128}}].

\bibitem{Heisenberg:2023prj}
L.~Heisenberg, N.~Yunes and J.~Zosso, \emph{{Gravitational wave memory beyond
  general relativity}},
  \href{https://doi.org/10.1103/PhysRevD.108.024010}{\emph{Phys. Rev. D}
  {\bfseries 108} (2023) 024010}
  [\href{https://arxiv.org/abs/2303.02021}{{\ttfamily 2303.02021}}].

\bibitem{Ohashi:2004wr}
A.~Ohashi and M.-a.~Sakagami, \emph{{Massive quasi-normal mode}},
  \href{https://doi.org/10.1088/0264-9381/21/16/010}{\emph{Class. Quant. Grav.}
  {\bfseries 21} (2004) 3973}
  [\href{https://arxiv.org/abs/gr-qc/0407009}{{\ttfamily gr-qc/0407009}}].

\bibitem{Konoplya:2004wg}
R.A.~Konoplya and A.V.~Zhidenko, \emph{{Decay of massive scalar field in a
  Schwarzschild background}},
  \href{https://doi.org/10.1016/j.physletb.2005.01.078}{\emph{Phys. Lett. B}
  {\bfseries 609} (2005) 377}
  [\href{https://arxiv.org/abs/gr-qc/0411059}{{\ttfamily gr-qc/0411059}}].

\bibitem{Konoplya:2006br}
R.A.~Konoplya and A.~Zhidenko, \emph{{Stability and quasinormal modes of the
  massive scalar field around Kerr black holes}},
  \href{https://doi.org/10.1103/PhysRevD.73.124040}{\emph{Phys. Rev. D}
  {\bfseries 73} (2006) 124040}
  [\href{https://arxiv.org/abs/gr-qc/0605013}{{\ttfamily gr-qc/0605013}}].

\bibitem{Zhidenko:2006rs}
A.~Zhidenko, \emph{{Massive scalar field quasi-normal modes of higher
  dimensional black holes}},
  \href{https://doi.org/10.1103/PhysRevD.74.064017}{\emph{Phys. Rev. D}
  {\bfseries 74} (2006) 064017}
  [\href{https://arxiv.org/abs/gr-qc/0607133}{{\ttfamily gr-qc/0607133}}].

\bibitem{Bolokhov:2023bwm}
S.V.~Bolokhov, \emph{{Long-lived quasinormal modes and
  overtones\textquoteright{} behavior of holonomy-corrected black holes}},
  \href{https://doi.org/10.1103/PhysRevD.110.024010}{\emph{Phys. Rev. D}
  {\bfseries 110} (2024) 024010}
  [\href{https://arxiv.org/abs/2311.05503}{{\ttfamily 2311.05503}}].

\bibitem{Bolokhov:2023ruj}
S.V.~Bolokhov, \emph{{Long-lived quasinormal modes and oscillatory tails of the
  Bardeen spacetime}},
  \href{https://doi.org/10.1103/PhysRevD.109.064017}{\emph{Phys. Rev. D}
  {\bfseries 109} (2024) 064017}.

\bibitem{Lutfuoglu:2025hwh}
B.C.~Lütfüoğlu, \emph{{Long-lived Quasinormal modes around regular black
  holes and wormholes in Covariant Effective Quantum Gravity}},
  \href{https://arxiv.org/abs/2504.09323}{{\ttfamily 2504.09323}}.

\bibitem{Dubinsky:2025bvf}
A.~Dubinsky, \emph{{Long-Lived Quasinormal Modes and Quasi-Resonances around
  Non-Minimal Einstein-Yang-Mills Black Holes}},
  \href{https://arxiv.org/abs/2505.08545}{{\ttfamily 2505.08545}}.

\bibitem{Konoplya:2011qq}
R.A.~Konoplya and A.~Zhidenko, \emph{{Quasinormal modes of black holes: From
  astrophysics to string theory}},
  \href{https://doi.org/10.1103/RevModPhys.83.793}{\emph{Rev. Mod. Phys.}
  {\bfseries 83} (2011) 793} [\href{https://arxiv.org/abs/1102.4014}{{\ttfamily
  1102.4014}}].

\bibitem{LIGOScientific:2014pky}
{\scshape LIGO Scientific} collaboration, \emph{{Advanced LIGO}},
  \href{https://doi.org/10.1088/0264-9381/32/7/074001}{\emph{Class. Quant.
  Grav.} {\bfseries 32} (2015) 074001}
  [\href{https://arxiv.org/abs/1411.4547}{{\ttfamily 1411.4547}}].

\bibitem{LISAConsortiumWaveformWorkingGroup:2023arg}
{\scshape LISA Consortium Waveform Working Group} collaboration,
  \emph{{Waveform Modelling for the Laser Interferometer Space Antenna}},
  \href{https://arxiv.org/abs/2311.01300}{{\ttfamily 2311.01300}}.

\bibitem{Degollado:2014vsa}
J.C.~Degollado and C.A.R.~Herdeiro, \emph{{Wiggly tails: a gravitational wave
  signature of massive fields around black holes}},
  \href{https://doi.org/10.1103/PhysRevD.90.065019}{\emph{Phys. Rev. D}
  {\bfseries 90} (2014) 065019}
  [\href{https://arxiv.org/abs/1408.2589}{{\ttfamily 1408.2589}}].

\bibitem{Damour:2007ap}
T.~Damour and S.N.~Solodukhin, \emph{{Wormholes as black hole foils}},
  \href{https://doi.org/10.1103/PhysRevD.76.024016}{\emph{Phys. Rev. D}
  {\bfseries 76} (2007) 024016}
  [\href{https://arxiv.org/abs/0704.2667}{{\ttfamily 0704.2667}}].

\bibitem{Gundlach:1993tp}
C.~Gundlach, R.H.~Price and J.~Pullin, \emph{{Late time behavior of stellar
  collapse and explosions: 1. Linearized perturbations}},
  \href{https://doi.org/10.1103/PhysRevD.49.883}{\emph{Phys. Rev. D} {\bfseries
  49} (1994) 883} [\href{https://arxiv.org/abs/gr-qc/9307009}{{\ttfamily
  gr-qc/9307009}}].

\bibitem{Konoplya:2024ptj}
R.A.~Konoplya, \emph{{Two regimes of asymptotic fall-off of a massive scalar
  field in the Schwarzschild\textendash{}de Sitter spacetime}},
  \href{https://doi.org/10.1103/PhysRevD.109.104018}{\emph{Phys. Rev. D}
  {\bfseries 109} (2024) 104018}
  [\href{https://arxiv.org/abs/2401.17106}{{\ttfamily 2401.17106}}].

\bibitem{Dubinsky:2024hmn}
A.~Dubinsky and A.~Zinhailo, \emph{{Asymptotic decay and quasinormal
  frequencies of scalar and Dirac fields around dilaton-de Sitter black
  holes}}, \href{https://doi.org/10.1140/epjc/s10052-024-13206-6}{\emph{Eur.
  Phys. J. C} {\bfseries 84} (2024) 847}
  [\href{https://arxiv.org/abs/2404.01834}{{\ttfamily 2404.01834}}].

\bibitem{Dubinsky:2024jqi}
A.~Dubinsky, \emph{{Telling late-time tails for a massive scalar field in the
  background of brane-localized black holes}},
  \href{https://doi.org/10.1209/0295-5075/ad51a3}{\emph{EPL} {\bfseries 147}
  (2024) 19003} [\href{https://arxiv.org/abs/2403.01883}{{\ttfamily
  2403.01883}}].

\bibitem{Konoplya:2020bxa}
R.A.~Konoplya and A.F.~Zinhailo, \emph{{Quasinormal modes, stability and
  shadows of a black hole in the 4D
  Einstein\textendash{}Gauss\textendash{}Bonnet gravity}},
  \href{https://doi.org/10.1140/epjc/s10052-020-08639-8}{\emph{Eur. Phys. J. C}
  {\bfseries 80} (2020) 1049}
  [\href{https://arxiv.org/abs/2003.01188}{{\ttfamily 2003.01188}}].

\bibitem{Skvortsova:2024atk}
M.~Skvortsova, \emph{{Quasinormal Frequencies of Fields with Various Spin in
  the Quantum Oppenheimer\textendash{}Snyder Model of Black Holes}},
  \href{https://doi.org/10.1002/prop.202400132}{\emph{Fortsch. Phys.}
  {\bfseries 72} (2024) 2400132}
  [\href{https://arxiv.org/abs/2405.06390}{{\ttfamily 2405.06390}}].

\bibitem{Bolokhov:2023dxq}
S.V.~Bolokhov, \emph{{Black holes in Starobinsky-Bel-Robinson Gravity and the
  breakdown of quasinormal modes/null geodesics correspondence}},
  \href{https://doi.org/10.1016/j.physletb.2024.138879}{\emph{Phys. Lett. B}
  {\bfseries 856} (2024) 138879}
  [\href{https://arxiv.org/abs/2310.12326}{{\ttfamily 2310.12326}}].

\bibitem{Malik:2024tuf}
Z.~Malik, \emph{{Analytical QNMs of fields of various spin in the Hayward
  spacetime}}, \href{https://doi.org/10.1209/0295-5075/ad7885}{\emph{EPL}
  {\bfseries 147} (2024) 69001}
  [\href{https://arxiv.org/abs/2410.04306}{{\ttfamily 2410.04306}}].

\bibitem{Malik:2024sxv}
Z.~Malik, \emph{{Quasinormal Modes of Dilaton Black Holes: Analytic
  Approximations}},
  \href{https://doi.org/10.1007/s10773-024-05660-5}{\emph{Int. J. Theor. Phys.}
  {\bfseries 63} (2024) 128}
  [\href{https://arxiv.org/abs/2409.09872}{{\ttfamily 2409.09872}}].

\bibitem{Konoplya:2024lch}
R.A.~Konoplya and O.S.~Stashko, \emph{{Probing the Effective Quantum Gravity
  via Quasinormal Modes and Shadows of Black Holes}},
  \href{https://arxiv.org/abs/2408.02578}{{\ttfamily 2408.02578}}.

\bibitem{Koyama:2001ee}
H.~Koyama and A.~Tomimatsu, \emph{{Asymptotic tails of massive scalar fields in
  Schwarzschild background}},
  \href{https://doi.org/10.1103/PhysRevD.64.044014}{\emph{Phys. Rev. D}
  {\bfseries 64} (2001) 044014}
  [\href{https://arxiv.org/abs/gr-qc/0103086}{{\ttfamily gr-qc/0103086}}].

\bibitem{Koyama:2001qw}
H.~Koyama and A.~Tomimatsu, \emph{{Slowly decaying tails of massive scalar
  fields in spherically symmetric space-times}},
  \href{https://doi.org/10.1103/PhysRevD.65.084031}{\emph{Phys. Rev. D}
  {\bfseries 65} (2002) 084031}
  [\href{https://arxiv.org/abs/gr-qc/0112075}{{\ttfamily gr-qc/0112075}}].

\bibitem{Burko:2004jn}
L.~Burko and G.~Khanna, \emph{Universality of massive scalar field late time
  tails in black hole spacetimes},
  \href{https://doi.org/10.1103/PhysRevD.70.044018}{\emph{Phys. Rev. D}
  {\bfseries 70} (2004) 044018}.

\bibitem{Cardoso:2016rao}
V.~Cardoso, E.~Franzin and P.~Pani, \emph{{Is the gravitational-wave ringdown a
  probe of the event horizon?}},
  \href{https://doi.org/10.1103/PhysRevLett.116.171101}{\emph{Phys. Rev. Lett.}
  {\bfseries 116} (2016) 171101}
  [\href{https://arxiv.org/abs/1602.07309}{{\ttfamily 1602.07309}}].

\bibitem{Witek:2012tr}
H.~Witek, V.~Cardoso, A.~Ishibashi and U.~Sperhake, \emph{{Superradiant
  instabilities in astrophysical systems}},
  \href{https://doi.org/10.1103/PhysRevD.87.043513}{\emph{Phys. Rev. D}
  {\bfseries 87} (2013) 043513}
  [\href{https://arxiv.org/abs/1212.0551}{{\ttfamily 1212.0551}}].

\bibitem{Konoplya:2022pbc}
R.A.~Konoplya and A.~Zhidenko, \emph{{First few overtones probe the event
  horizon geometry}},
  \href{https://doi.org/10.1016/j.jheap.2024.10.015}{\emph{JHEAp} {\bfseries
  44} (2024) 419} [\href{https://arxiv.org/abs/2209.00679}{{\ttfamily
  2209.00679}}].

\bibitem{Konoplya:2023hqb}
R.A.~Konoplya, \emph{{The sound of the event horizon}},
  \href{https://doi.org/10.1142/S0218271823420142}{\emph{Int. J. Mod. Phys. D}
  {\bfseries 32} (2023) 2342014}
  [\href{https://arxiv.org/abs/2312.16249}{{\ttfamily 2312.16249}}].

\bibitem{Ishihara:2005dp}
H.~Ishihara and K.~Matsuno, \emph{{Kaluza-Klein black holes with squashed
  horizons}}, \href{https://doi.org/10.1143/PTP.116.417}{\emph{Prog. Theor.
  Phys.} {\bfseries 116} (2006) 417}
  [\href{https://arxiv.org/abs/hep-th/0510094}{{\ttfamily hep-th/0510094}}].

\bibitem{Gregory:1993vy}
R.~Gregory and R.~Laflamme, \emph{{Black strings and p-branes are unstable}},
  \href{https://doi.org/10.1103/PhysRevLett.70.2837}{\emph{Phys. Rev. Lett.}
  {\bfseries 70} (1993) 2837}
  [\href{https://arxiv.org/abs/hep-th/9301052}{{\ttfamily hep-th/9301052}}].

\bibitem{Kimura:2007cr}
M.~Kimura, K.~Murata, H.~Ishihara and J.~Soda, \emph{{Stability of Squashed
  Kaluza-Klein Black Holes}},
  \href{https://doi.org/10.1103/PhysRevD.77.064015}{\emph{Phys. Rev. D}
  {\bfseries 77} (2008) 064015}
  [\href{https://arxiv.org/abs/0712.4202}{{\ttfamily 0712.4202}}].

\bibitem{Bolokhov:2024voa}
S.V.~Bolokhov and R.A.~Konoplya, \emph{{Circumventing quantum gravity: Black
  holes evaporating into macroscopic wormholes}},
  \href{https://doi.org/10.1103/PhysRevD.111.064007}{\emph{Phys. Rev. D}
  {\bfseries 111} (2025) 064007}
  [\href{https://arxiv.org/abs/2410.10419}{{\ttfamily 2410.10419}}].

\bibitem{Konoplya:2025nqv}
R.A.~Konoplya, Z.~Stuchlík and A.~Zhidenko, \emph{{Charged black hole
  surrounded by a galactic halo in a de Sitter universe}},
  \href{https://doi.org/10.1103/l33g-6hlr}{\emph{Phys. Rev. D} {\bfseries 112}
  (2025) 083014} [\href{https://arxiv.org/abs/2509.03301}{{\ttfamily
  2509.03301}}].

\bibitem{Konoplya:2022hbl}
R.A.~Konoplya and A.~Zhidenko, \emph{{Solutions of the Einstein Equations for a
  Black Hole Surrounded by a Galactic Halo}},
  \href{https://doi.org/10.3847/1538-4357/ac76bc}{\emph{Astrophys. J.}
  {\bfseries 933} (2022) 166}
  [\href{https://arxiv.org/abs/2202.02205}{{\ttfamily 2202.02205}}].

\end{thebibliography}\endgroup

\end{document}